\documentclass[onecolumn,preprintnumbers,amsmath,amssymb]{revtex4}

\usepackage{amsmath}
\usepackage[dvips]{graphicx}
\usepackage{algorithm}
\usepackage{algorithmic}
\begin{document}

\title{Revealing Dynamic Communities in networks using genetic algorithm with Merging and Splitting Operators}

\author{Weihua Zhan$^{1,2}$}
\email{zhanweihua@zwu.edu.cn}

\author{Lei Deng$^{3}$}

\author{Jihong Guan$^{4}$}

\author{Jun Niu$^{5}$}

\affiliation{$^{1}$College of Control Science and Engineering, Zhejiang University, 38 Zheda Road, Hangzhou 310058,
China}
\affiliation{$^{2}$School of Electronics and Computer Science, Zhejiang Wanli University, 8 Qianhu Road, Ningbo 315100, China}
\affiliation{$^{3}$School of Software, Central South University, Changsha 410075, China}

\affiliation{$^{4}$Department of Computer Science and Technology, Tongji University, 4800 Cao'an Road, Shanghai 201804,
China}

\affiliation{$^{5}$Department of Computer Science and Technology, Ningbo University, 8 Feng'Hua Road, Ningbo 315211,
China}

\begin{abstract}
Community structure is pervasive in various real-world networks, portraying the strong local clustering of nodes. Unveiling the community structure of a network is deemed to a crucial step towards understanding the dynamics on the network. Actually, most of the real-world networks are dynamic and their community structures are evolutionary over time accordingly. How to revealing the dynamical communities has recently become a pressing issue. Here, we present an evolutionary method for accurately identifying dynamical communities in the networks.
In this method, we first introduced a fitness function that is a compound of asymptotic surprise values on the current and previous snapshots of the network. Second, we developed ad hoc merging and splitting operators, which allows for large-scale searching while preserving low cost. Third, this large-scale searching coupled with local mutation and crossover enhanced revealing a better solution to each snapshot of the network. This method does not require specifying the number of communities advanced, and free from resolution limit while satisfying temporal smooth constraint. Experimental results on both model and real dynamic networks show that the method can find a better solution compared with state-of-art approaches.
\end{abstract}

%\pacs{87.23.Ge, %ecology and evolution
%02.50.Le, %Game theory
%89.75.Fb %Self-organization complex systems
%}

\maketitle

%%%%%%%%%%%%%%%%%%%%%%%%%%%%%%%%%%%%%%%%%%%%%%%%%%%%%%%%%%%%%%%%%%%
%%%%%%%%%%%%%%%          1 Introduction           %%%%%%%%%%%%%%%%%
%%%%%%%%%%%%%%%%%%%%%%%%%%%%%%%%%%%%%%%%%%%%%%%%%%%%%%%%%%%%%%%%%%%
\section{INTRODUCTION\protect}
{N}{etworks} or graphs are widely used to model various complex systems, and network analysis has found many applications in diverse contexts~\cite{Erkan04,veronis04,Lv02,Zhang17,Xiao17}. It has been discovered that many real-world networks possess some common structure characteristics~\cite{Watts98,Barabasi99,Newman03}. An interesting characteristic is community structure\cite{Girvan:2002a}, i.e., the nodes in a network can be clustered in groups, with dense internal connection but sparse external connections. A number of approaches to identify community structure have been presented for static networks, such as modularity optimization~\cite{Newman06a,Newman13,Guimera05,Duch05}, propagation~\cite{Raghavan07},non-negative matrix fraction~\cite{Zhang07a}, information theory methods~\cite{Rosvall07,Rosvall08}, and inference-based methods\cite{Hasting06,Newman07}(see the literature~\cite{Fortunato10} for a detail review). Nevertheless, real-world networks almost always vary, and their community structures evolve over time accordingly. For instance, sometimes a researcher in a scientist coauthor network may coauthor a paper with one never worked together before, which could indicate that she~(or he) is moving towards a new academic circle.

In essence, a dynamic network consists of a series of snapshots at different time steps. A plain strategy for coping with dynamic networks is first to detect communities on their snapshot networks separately using static methods. And tracking community evolution can thus be achieved by successively matching community structures between adjacent snapshots. There are two disadvantages, however, in such methods. Firstly, network data is gathered with some noises. Secondly, a given network has many good partitions which compete with each other while disagree on the partition structure~\cite{Good10}. This will cause the couples of matched communities endure abrupt changes, which is undesirable in the practical context.

To capture the real evolution of community structure, a better alternative is to use a so-called evolutionary clustering framework that was first proposed by Chakrabarti et al.~\cite{Chakrabarti06}. This framework assumes that clustering should consider temporal smoothness between adjacent snapshots. Specifically, community structure at
time $t$ is influenced by the clusters at time $t-1$. Recently, this framework has become popular in the study of dynamical community detection~\cite{lin09,Chi09,Kim09,Folino14,Qin16,Wang17}.

Lin et al.~\cite{lin09} proposed FacetNet framework for systemic analyzing communities and their evolution, which uses the already identified community structure at time $t-1$ to regularize the community structure at current time $t$. The cost function to optimize was thus defined as the combination of the snapshot cost and temporal cost. To identify community structure at each snapshot, the authors presented a method based on non-negative matrix factorization to optimize the cost function.
 Incorporating the temporal smoothness in spectral clustering, Chi et al.~\cite{Chi09} presented two frameworks for evolutionary spectral clustering, PCQ~(Preserving Cluster Quality ) and PCM~(Preserving Cluster Membership). Qin et al.~\cite{Qin16}proposed a multi-similarity spectral clustering (MSSC) method and a dynamic co-training algorithm for community detection in dynamic networks. This method preserves the evolutionary information of community structure by combining the current data and historic partitions. Wang et al~\cite{Wang17}defined a new similarity by combing structural perturbation and topological features. Based on this similarity, they defined a cost function incorporating temporal smoothness to be optimized for dynamic community detection. An practical limitation for most of these method is that it is required specifying the number of communities $k$ in advance. For instance, Lin et al.~\cite{lin09} suggested to execute FacetNet with different $k$, and then select the partition with the highest modularity as the best one.

 Another paradigm for considering temporal smoothness is to formulate it as one objective to optimize. Dynamic community detection can thus be formulated as a multiobjective optimization problem, in which temporal cost is given by the mutual information between the partitions on adjacent snapshots. Gong et al.~\cite{Gong12} proposed a multiobjective immune algorithm, where  simultaneously optimize the modularity and normalized mutual information. Folino et al.~\cite{Folino14} presented an evolutionary multiobjective approach, DYNMOGA, for dynamic community detection. The work also used the mutual information as the temporal cost, while the snapshot cost is measured by four quality functions including modularity, conductance~\cite{Kannan04}, normalized cut~\cite{Shi00}, and community score~\cite{Folino10}. To get a single solution out of the Pareo front, DYNMOGA selects the partition with the highest modularity.

As we can see, these evolutionary clustering methods rely on a temporal cost function to be optimized. Hence, the accuracy of an evolutionary clustering method depends on two crucial points. First of all, a temporal cost function is desirable which can accurately measuring quality of a partition on snapshots. The second, an efficient and accurate method need to be selected or developed for optimization of the cost function.

Here, we present an evolutionary approach for dynamic community detection by employing a genetic algorithm, named $\textbf{MSGA}$~(Genetic Algorithm with Merging and Splitting operators). Based on the asymptotical surprise~\cite{Traag15} and temporal smoothness constraint, we introduced asymptotical surprise that can accurately measure the quality of partitions on the dynamic network and be quickly calculated. The dynamic community detection thus is formulated as a problem of maximizing the temporal asymptotic surprise. To efficiently optimize the cost function, we developed ad hoc split operator and merge operator with a low time cost. Using locus-based adjacency encoding, this approach  has an advantage that it does not require specifying an assumed number of communities. Moreover, the method is free of resolution limit which makes it can reveal community structure accurately, compared with state-of-art approaches.

\section{Proposed Method}
\subsection{Fitness function}
The most popular measure for the evaluating the quality of a partition on a static network is modularity~\cite{Newman04a}. However, it is not a good choice to extend this measure to the dynamic network, since this measure has an inherent resolution limit~\cite{Fortunato07}. To confront with the resolution limit, some alternative measures have been developed, such as modular density~\cite{Li08}, absolute Potts model~\cite{Ronhovde10} and Surprise\cite{Aldecoa11}. It has been shown that Surprise can break the resolution limit~\cite{Nicolini16} and surprise maximization can be used to precisely reveal the community structure of networks~\cite{Aldecoa13}. However, optimizing Surprise is a difficult task due to the high computation complexity. To our best knowledge, FASGO~\cite{Jiang14} is so far the unique algorithm to optimize Surprise, which is devised using greedy strategy.

Recently, Traag et al~\cite{Traag15} proposed an accurate asymptotical formulation of Surprise, called as asymptotical Surprise~(AS). Let the number of nodes in the network of interest be $n$, and the number of links be $m$. The total number of possible links is $M=\frac{n\times{(n-1)}}{2}$. Given a partition $\pi$, we can obtain the total numbers of internal links $m_{int}$ and possible internal links $M_{int}$. The asymptotical Surprise
\begin{equation}
\centering
\label{AS}
AS(\pi,G)=mD(q||<q>).
\end{equation}
where $q$ is the ratio of total internal links $q=\frac{m_{int}}{m}$, and $<q>$ is the ratio of total possible internal links, and $D(x||y)$ is the KL divergence. It is easily to find that AS can be quickly calculated with complexity $O(m)$.

To deal with dynamic community detection, we extend AS to temporal asymptotical Surprise~(\textbf{TAS}). A dynamic network $\mathbb{G}$ is given by a series of network snapshots, $\mathbb{G}=\{G_1,G_2,\ldots,G_T\}$. For a given partition $\pi$, temporal asymptotical Surprise at time step $t$ is defined as Eq.~(\ref{eq:TASfull}), where $1\le k\le t-1$.

\begin{equation*}\label{eq:TASfull}
TAS(\pi,\mathbb{G},t)=
\begin{cases}
AS(\pi,G_1),\qquad t=1; \\
\sum_{i=1}^{k}(1-\beta)^{i}\cdot AS(\pi,G_{t-i})+\beta\cdot AS(\pi,G_t), t=2,3,\ldots,T
\end{cases}
\end{equation*}

\begin{equation*}\label{eq:tasreduced}
TAS(\pi,\mathbb{G},t)=\\
\begin{cases}
AS(\pi,G_1)  , \qquad t=1 \\
\beta\cdot AS(\pi,G_t)+(1-\beta)\cdot AS(\pi,G_{t-1}) , t=2,3,\ldots,T
\end{cases}
\end{equation*}

That is, temporal asymptotical Surprise at time step $t$ depends on the AS of $t$,$t-1$,$\ldots$, $t-k$. Furthermore, the more distant from current time step, the less contribution to the current TAS is. To effectively calculate TAS, $k$ takes the value 1, and TAS is then reduced to Eq.~(\ref{eq:TASreduced}).

In MSGA, the reduced version of TAS is used as fitness function of a chromosome with complexity $O(m)$. Let $\Delta^+$ denote the set of new established links against previous time step, and $\Delta^-$ denote the set of removed links. When the links difference set($\Delta^+,\Delta^-$) between $G_t$ and $G_{t-1}$ has been figured out, the reduced TAS can be further speedup.

\subsection{Genetic algorithm with merging and splitting operators}

As opposed to other heuristics, evolutionary approaches have a stronger ability of global search originated from the search mechanism based on population, which allow us to find a better solution. it was recently witnessed that evolutionary methods succeed in the community detection problem~\cite{Pizzuti08,zhan11,zhan16}. Here, we presented an evolutionary approach for dynamic community detection, MSGA, which is based on the elite genetic algorithm, and equipped with ad hoc merging and splitting operators. As DYNMOGA and other evolutionary approaches for community detection, MSGA uses locus-based adjacency encoding schema. This schema brings to these approaches an advantage that they do not require specifying the number of communities in advance. Under the encoding schema, a community in a partition encoded by a chromosome is represented as an approximate span tree.

Merging operator consists of two steps: selecting two groups of nodes and performing merging. The first group is randomly picked out, and the second is the group that has the high link density to the first group. If the total degree of nodes in the first group is $D_1$, then the complexity to select the second group is $O(D_1)$. To perform merging the two groups amounts to construct a spanning tree likewise that encompasses all the nodes in the two groups by modifying the loci associated these nodes. Let $D_2$ be the total degree of the nodes in the second group. The complexity of performing merging is $O(D_1+D_2)$ since it involves traversing all the links of these nodes. Hence, the total complexity of merging operator is $O(D_1+D_2)$.

The splitting operator is to divide a group into two subgroups. This operator is also composed of two steps: selecting a group of nodes to split randomly and performing splitting. We can randomly pick a node $i$ out, and select the group $C_i$ to be split which node $i$ is assigned to in the partition encoded by a chromosome. In this way, the first step can be implemented in time O(1). To split the group, we can use spectral bipartition method~\cite{Newman06a} to obtain two subgroups and then construct two approximate span trees to encode the two subgroups. It can be verified the overall complexity is $O(\left|C_i\right|)$.

Algorithm~\ref{alg:MSGA} describes the overall process of the algorithm for dynamic community detection.
%\begin{algorithm}
%\caption{Fitness for dynamic community detection}
%\label{alg:extractHierarchy}
%\renewcommand{\algorithmiccomment}[1]{// #1}
%\renewcommand{\algorithmicrequire}{ \textbf{Input:}}      %Use Input in the format of Algorithm
%\renewcommand{\algorithmicensure}{ \textbf{Output:}}
%\begin{algorithmic}[1]
%\REQUIRE $G_t$,$G_{t-1}$,$\pi$,~$\Delta^{+}$,$\Delta^{-}$\\
%\ENSURE TAS($\pi$,$\mathbb{G}$,$t$) \\
%\STATE SET $ p=0$ ,$M=0$, Initialize each element in $vsize$ with 0
%\FORALL{ $i$ in $G_t$}
%  \STATE  $k\leftarrow \pi(i)$
%  \STATE  $vsize[k]\leftarrow vsize[k]+1$;
%\ENDFOR
%\FORALL{ $i$ in $G_t$}
%\FOR{ $j$ in Neighbor(i)}
%\IF{$\pi(i)=\pi(j)$} \STATE\COMMENT{node $i$ and $j$ are from the same community}
%\STATE $p_t\leftarrow p_t+1$
%\ENDIF
%\ENDFOR
%\ENDFOR
%\FORALL {$k \in$ \{1,..$\left|\pi\right|$,\}}
%\STATE $M\leftarrow (vsize[k]\times vsize[k])/2$
%\ENDFOR
%\STATE AS($\pi$,$G_t$)$\leftarrow$ DL($\frac{p_t}{m_t}\parallel\frac{M}{F}$)
%\STATE $p_{t-1}\leftarrow p_t$
%\FORALL{link $e$ in $\Delta^{+}$}
%\IF{$\pi(Adj1(e))=\pi(Adj2(e)$}
%\STATE $p_{t-1}\leftarrow p_{t-1}-1$
%\ENDIF
%\ENDFOR
%\FORALL{link $e$ in $\Delta^{-}$}
%\IF{$\pi(Adj1(e))=\pi(Adj2(e)$}
%\STATE $p_{t-1}\leftarrow p_{t-1}+1$
%\ENDIF
%\ENDFOR
%\STATE AS($\pi$,$G_{t-1}$)$\leftarrow$ DL($\frac{p_{t-1}}{m_{t-1}}\parallel\frac{M}{F}$)
%\STATE TAS($\pi$,$\mathbb{G}$,$t$)$\leftarrow \beta\times$ AS($\pi$,$G_t$)+$(1-\beta)\times$ AS($\pi$,$G_{t-1}$)
%\end{algorithmic}
%\end{algorithm}

\begin{algorithm}
\caption{MSGA~(Genetic algorithm with merge and split operation)}
\label{alg:MSGA}
\renewcommand{\algorithmiccomment}[1]{// #1}
\renewcommand{\algorithmicrequire}{ \textbf{Input:}}      %Use Input in the format of Algorithm
\renewcommand{\algorithmicensure}{ \textbf{Output:}}
\begin{algorithmic}[1]
\REQUIRE $\mathbb{G}$,$T$,$GEN$,$popuSize$,~$crossRate$\\
\ENSURE $\pi_1$,$\pi_2$,...,$\pi_T$ \\
\STATE SET $ t=1$
\WHILE{ $t\le T$ }
  \IF{$T=1$}
  \STATE  Initialize population with random individuals;
  \ELSE
  \STATE  Initialize population with part of members being of best individuals from $t-1$ generation;
  \STATE  Evaluate the set of links difference~(~$\Delta^{+}$,$\Delta^{-}$) between $G_t$ and $G_{t-1}$
  \ENDIF
  \STATE Evaluate the fitness of each individual in the population;
  \STATE i=2;
  \WHILE{$i\le GEN$}
      \STATE Top $10\%$ individuals in the previous population are duplicated to the current population
      \STATE j=0;
      \WHILE{$j<0.9\times popuSize$}
            \STATE Select two individuals $I_1$ and $I_2$ from the previous population
            \STATE $r\leftarrow rand()$
            \IF{$r\le crossRate$}
                 \STATE Perform crossover on the pair of individuals
            \ELSE
                \STATE Perform mutation on $I_1$ and $I_2$
                \STATE Randomly choose $I_1$ or $I_2$ to perform merging and splitting operations
             \ENDIF
             \STATE $j\leftarrow j+2$
      \ENDWHILE
      \STATE Evaluate the fitness of each individual in the population
      \STATE $i\leftarrow i+1$
  \ENDWHILE
  \STATE $t\leftarrow t+1$
\ENDWHILE
\end{algorithmic}
\end{algorithm}
This algorithm is feed with the dynamic network $\mathbb{G}$, the number of time steps $T$, population size $popuSize$, and $crossRate$. It outputs the best partition for each time step. The main operations for each snapshot
are listed as follows:

(1) Initiating population. For the first time step each chromosome are randomly generated in the initialization
population. However, for other time steps most of individuals are randomly generated and with the rest are best individuals duplicated from previous time step. Furthermore, it requires computing the difference set of links between the current and previous time steps.

(2) Calculating the fitness of each individual. We decoded a chromosome into a partition and then calculate the TAS associated the current partition using equation~\ref{eq:tasreduced}.

(3) Generating a new population. The 10\% of the population come from the previous time step. The 90\% individuals are reproduced in the following way: $crossRate$ individuals are generated with crossover operation, the remaining are obtained by performing mutation, merging and splitting operations.

The merging and splitting operator are important in the algorithm. Compared with mutation operation, these two operations perform large-scale searching as crossover but are more targeted, which facilitates to find a better solution.

\section{Experimental Results and Analysis}

An extensively used measure for the distance between an identified partition and the ground truth is normalized mutual information~(NMI)~\cite{Danon05}.  This measure takes its maximum value 1 when the
found division perfectly matches the ground truth, while it takes the minimum value 0 if they are totally independent of each other. Recently, error rate~\cite{lin09} is frequently coupled with NMI for evaluating the distance between an identified partition and the ground truth. Let the number of nodes in the network be $n$. If the number of communities in an identified partition is $k$, then we can use an $n\times k$ matrix $Z$ to indicate the partition. Likewise, we can use an $n\times m$ matrix $G$ to indicate the ground truth if the true number of communities is $m$. So the error rate is just the norm
$\left \lVert ZZ^T-GG^T \right \rVert$, which can be proved that the value equals to the number of the nonzero entries of $ZZ^T-GG^T$.

To evaluate the effectiveness of MSGA, we first test this algorithm on serval benchmarks of dynamic networks. We compare MSGA with DYNMOGA and FacetNet, which are two well-known evolutionary clustering approaches for dynamic community detection. As an evolutionary method, it requires to setting serval parameters advanced like DYNMOGA. We set crossover rate=0.7, mutation rate=0.7, population size=100 and number of generations is fixed 500.

\subsection*{Synthetic dataset \#1}

In order to evaluate the performance of an algorithm, Girvan and Newman ~\cite{Girvan:2002a} proposed a benchmark network that consists of $4$ communities of 32 nodes. Each node has a fixed average degree 16 and an adjustable number of external links $z$ that controls the strength of community structure of the network. Increasing $z$ the community structure becomes more fuzzy and the difficulty to recover true partitions increase until to a threshold that any algorithm cannot find meaningful modular structure. To introduce dynamics to a network, we can switch the memberships of $nC\%$ nodes at each time step. The instance network can be conveniently generated by the tools
provided by Greene~\cite{Greene10}.

\begin{figure}[ht]
\centering
\includegraphics[scale=0.22]{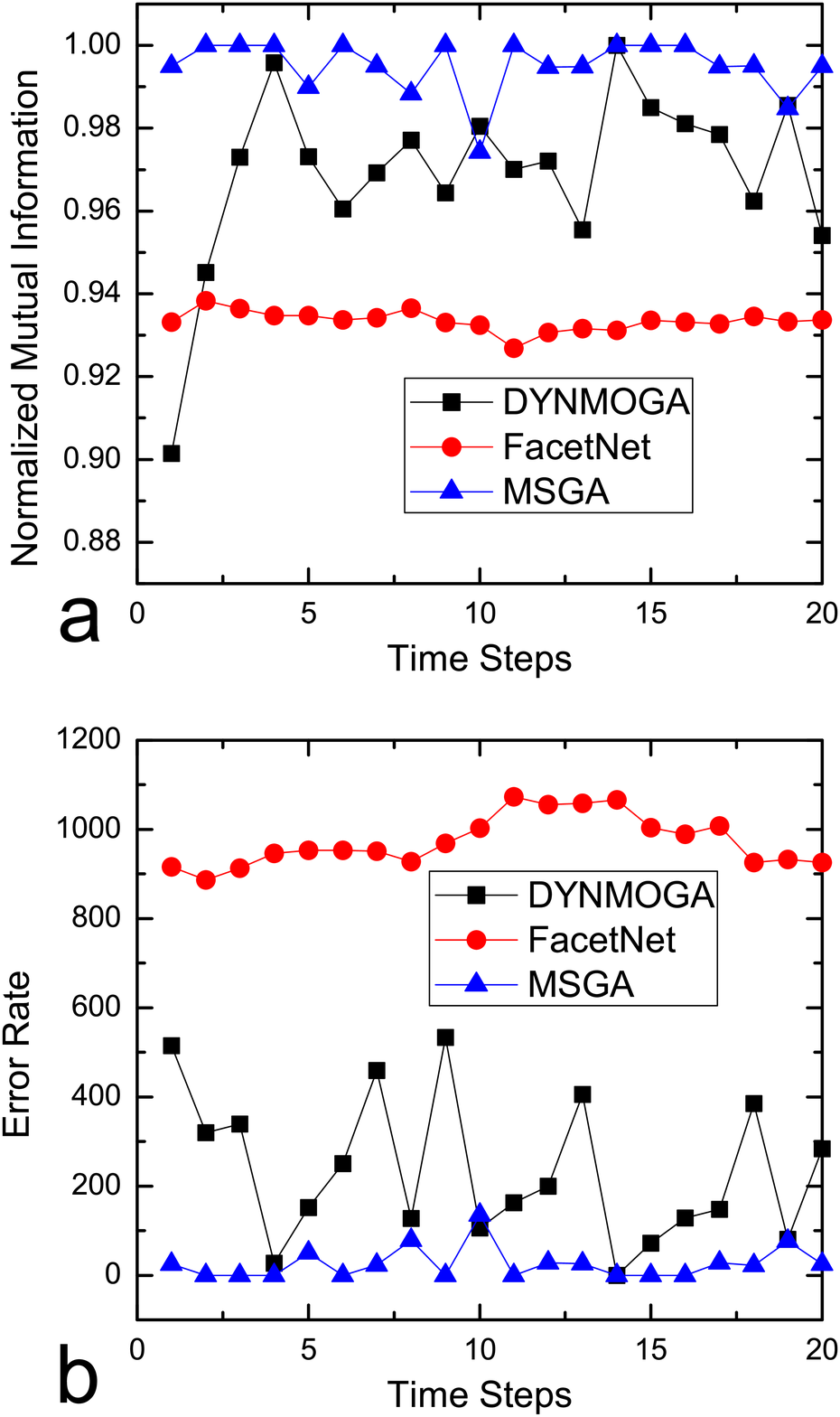}
\caption{Performances of the algorithms on switch network~(Z=5,P=10\%) over 20 time steps.}
\label{fig:switchminfoz5}
\end{figure}

Figure.~\ref{fig:switchminfoz5} shows the performance of these algorithms over twenty time steps. As we can see from Figure.~\ref{fig:switchminfoz5}~(a), any of these algorithms can find a good partition because at almost
all time steps the NMIs of all algorithms are greater than $0.9$. However, MSGA significantly outperforms FacetNet at all time steps and outperform DYNMOGA at most time steps. At 18 time steps, the NMIs of MSGA exceed 0.99 and are greater than those of DYNMOGA.  In contrast, only for time step 10 and 19 the NMIs of DYNMOGA are slightly greater than MSGA,  with the values of 0.9805 and 0.9855~(for MSGA, 0.9743 and 0.9848) while for other steps below 0.98. Figure.~\ref{fig:switchminfoz5}~(b) demonstrates the result of performance comparison in terms of error rate. As we can see, the result is in perfect accordance with that on NMI. MSGA exhibits a higher accuracy than DYNMOGA and FacetNet. Error rates of MSGA are less than 100 at all time steps except time step 10. At this step the error rate is 135, which is the unique time step slightly higher than that of DYNMOGA but evidently less than the maximum value of DYNMOGA 533 at time step 9. Furthermore, the error rates of DYNMOGA fluctuate relatively frequently and with relatively high values.

\begin{figure}[ht]
\centering
\includegraphics[scale=0.22]{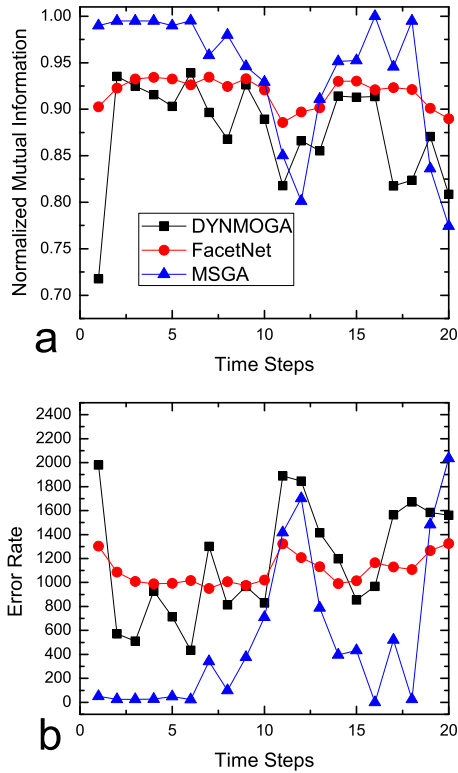}
\caption{Performances of the algorithms on switch network~(Z=6,P=10\%) over 20 time steps.}
\label{fig:switchz6p1}
\end{figure}

\begin{figure}[ht]
\centering
\includegraphics[scale=0.22]{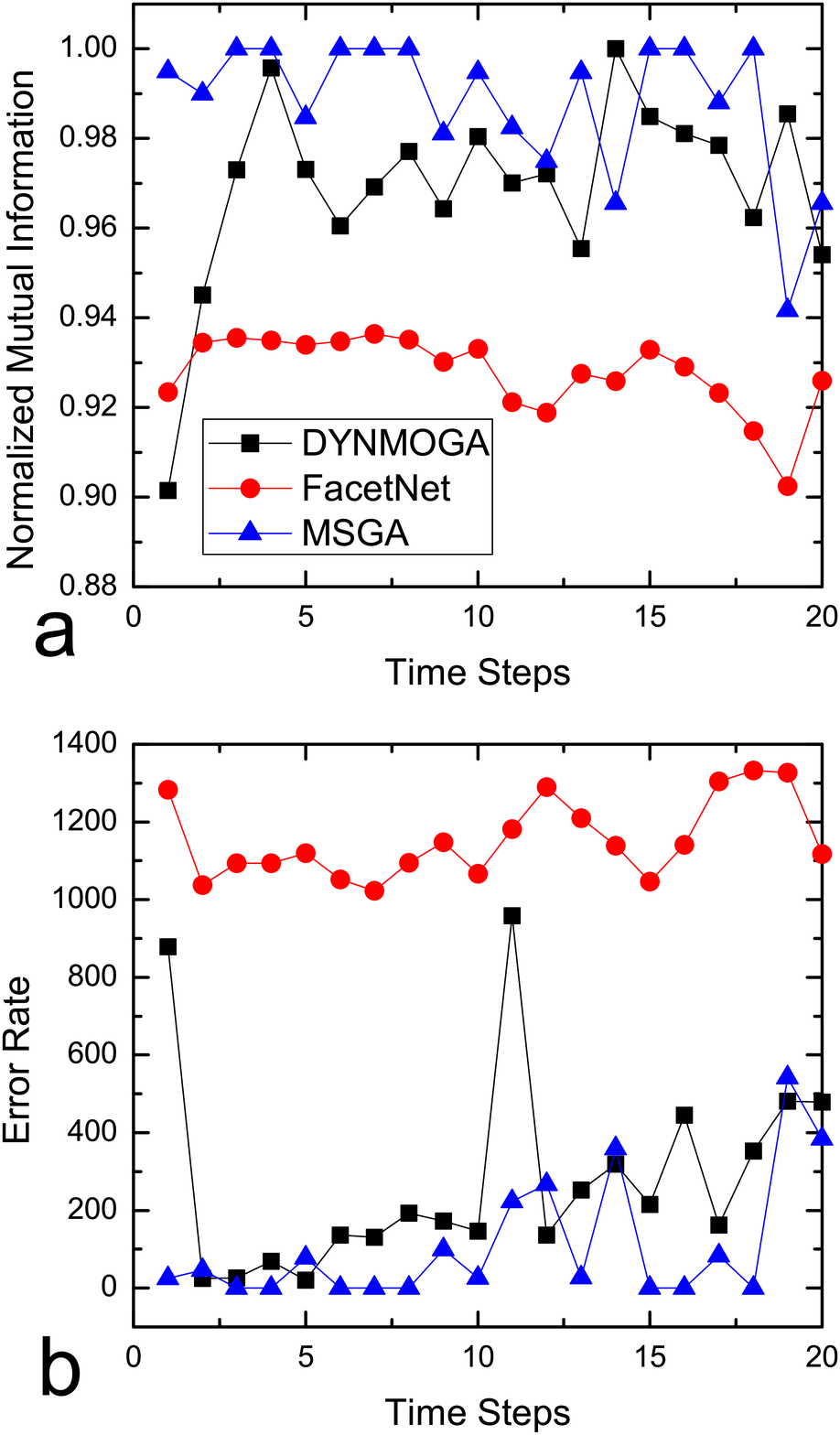}
\caption{Performances of the algorithms on switch network~(Z=5,P=30\%) over 20 time steps.}
\label{fig:switchz5p3}
\end{figure}

To be aware of the performance of MSGA on the network with other parameters, we checked two cases: (1) Z=6,p=10\% and (2) Z=5,p=30\%. Figure~\ref{fig:switchz6p1} and \ref{fig:switchz5p3} show the performance of these algorithm under the two cases.
\subsubsection*{Synthetic dataset \#2}

For real-world networks both the distributions of node degree and size of communities are observed to commonly followed power-law distribution~\cite{Barabasi99}. LFR benchmark~\cite{Lancichinetti08} is more close to real-world networks by imposing power-law distributions on node degrees and community sizes. Based on LFR benchmark, Greene et al~\cite{Greene10} developed a network generator that is able to generate dynamic network instances with power-law distributions and embedded community events. The embedded community events include:
\begin{itemize}
\item Birth and Death: starting from the second time step, 10\% of existing communities are randomly removed and 10\% of new communities are created by removing nodes from other communities.
\item Merging and splitting: starting from the second time step,10\% of communities are split, 10\% of communities are chosen and are pairwise merged.
\item Expansion and contraction: starting from the second time step, 10\% of communities are randomly chosen and then expanded or contracted by 25\% of their size. When expanded, the new nodes come from other communities.
\item Intermittent communities: starting from the second time step,10\% of communities disappear and then reappear at the next time step.
\end{itemize}

By Greene's tool, we generated with the same parameters four data sets combined the dynamic events respectively. Each network consists of 1000 nodes, with mean degree of 15, maximum degree 50 each and mixing coefficient fixed to 0.2. The power exponents for degree and community size are set to -1 and -2, respectively.

\begin{figure}[ht]
\centering
\includegraphics[scale=0.3]{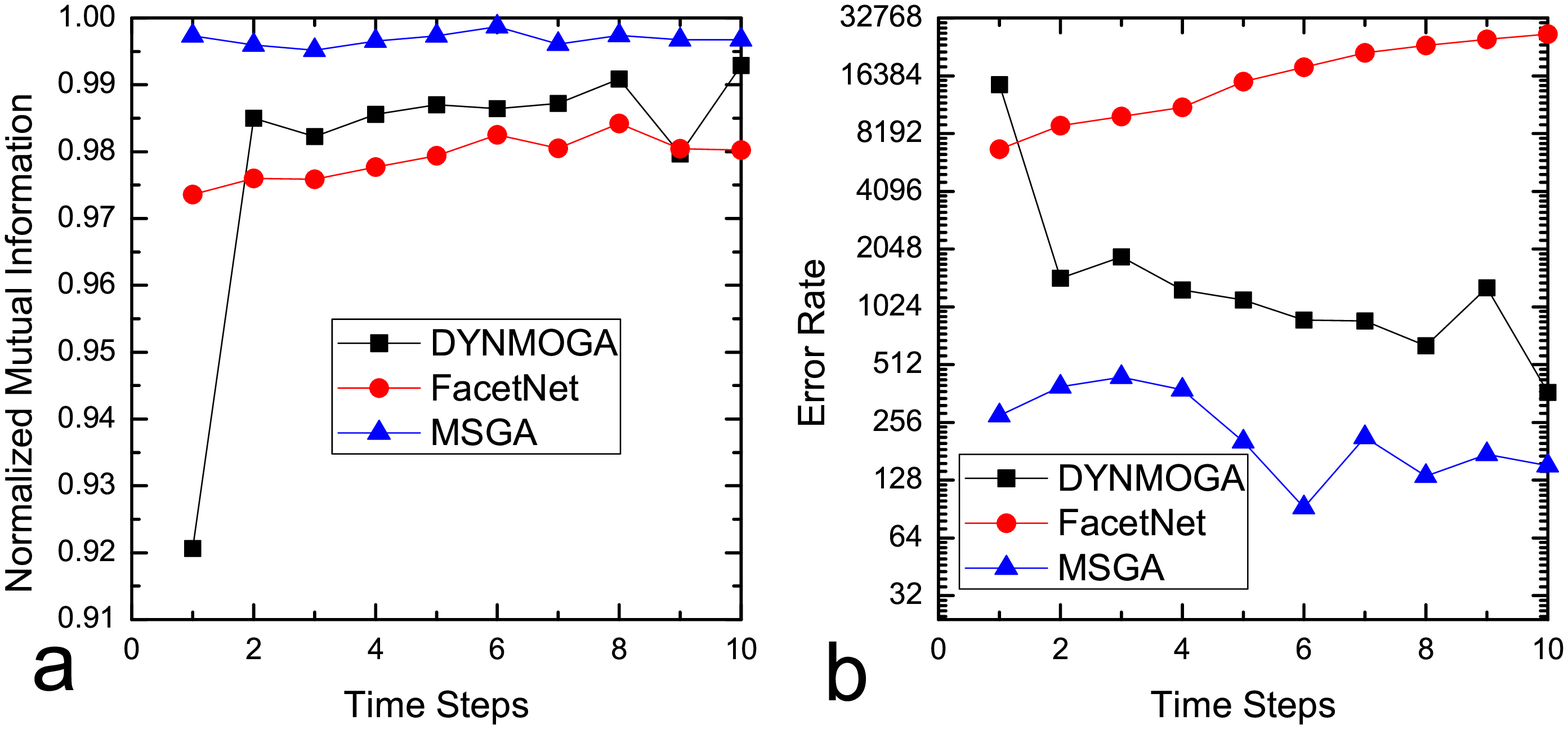}
\caption{Performances of the algorithms on birth and death network over 10 time steps.}
\label{fig:birthdeath}
\end{figure}

Figure~\ref{fig:birthdeath} illustrates the performance of these algorithms on the birth and death network.
As shown in the figure~\ref{fig:birthdeath}(a), MSGA exhibits a greater accuracy than DYNMOGA and FacetNet in terms of NMI. In figure~\ref{fig:birthdeath}(b) log-2 scaling of y axis is used to distinguish the performance between MSGA and DYNMOGA in term of error rate.

\begin{figure}[ht]
\centering
\includegraphics[scale=0.3]{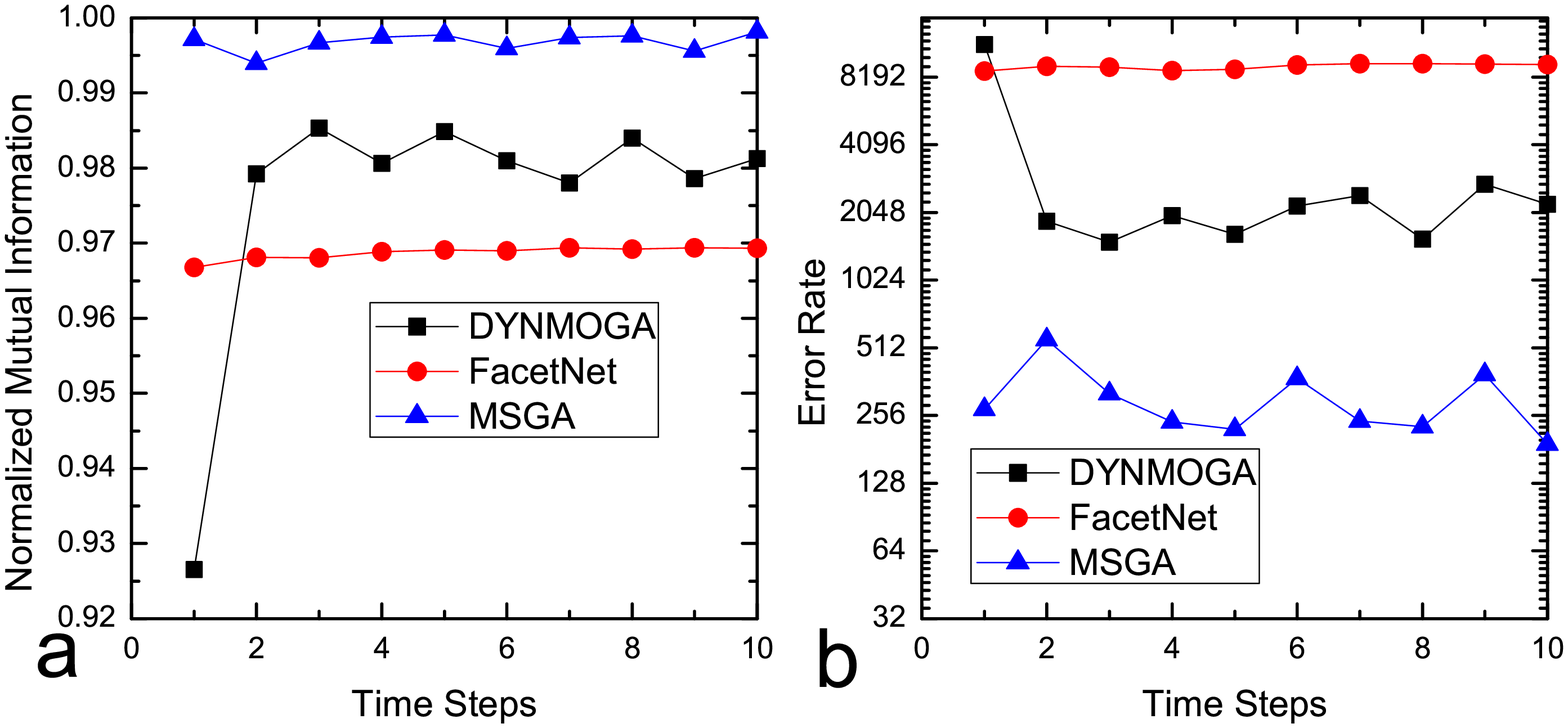}
\caption{Performances of the algorithms on expand and contraction network over 10 time steps.}
\label{fig:expand}
\end{figure}

Figure~\ref{fig:expand} shows the performance variations of these algorithms on expand and contraction network over 10 time steps. Overall, the performance of each algorithm is good and slightly fluctuates around own average value over all the time steps. At all the time steps MSGA has the best accuracy on the network in terms of both NMI and error rate. In terms of NMI, the values associated with MSGA are more close to 1 for any time step. As opposed to NMI, error rate looks more sensible to the performance difference among these algorithms. At any time step, MSGA is with the lowest error rate of around 300, DYNMOGA with around 2000, while FacetNet is with around 8000.

\begin{figure}[ht]
\centering
\includegraphics[scale=0.3]{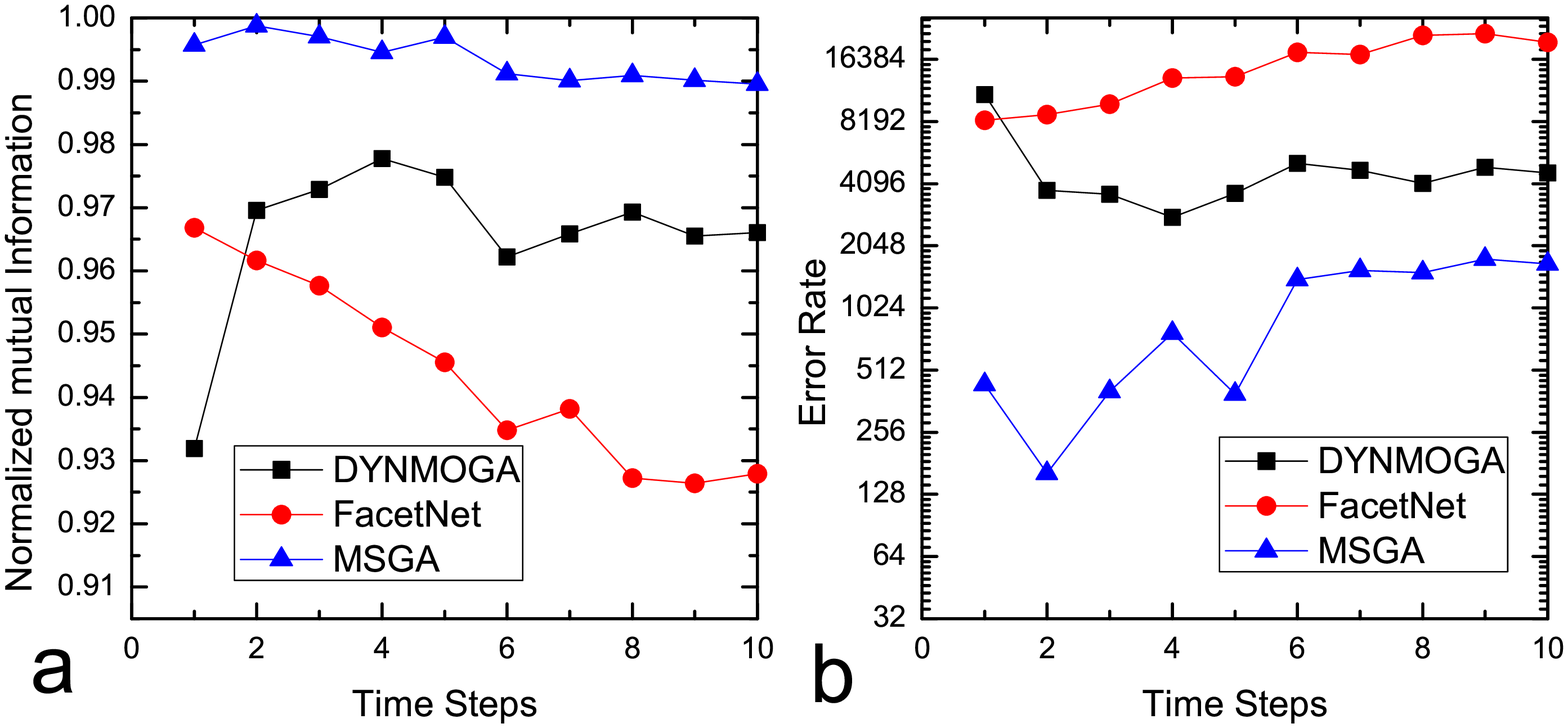}
\caption{Performances of the algorithms on merging and splitting network over 10 time steps.}
\label{fig:mergesplit}
\end{figure}

Figure~\ref{fig:mergesplit} illustrates the performance comparison among these algorithms on merge and split network. As a whole, the performance of FacetNet linearly degrade over the time steps in terms of both NMI and  error rate. In contrast, the performance curves of MSGA and DYNMOGA are relatively flat.

\begin{figure}[ht]
\centering
\includegraphics[scale=0.3]{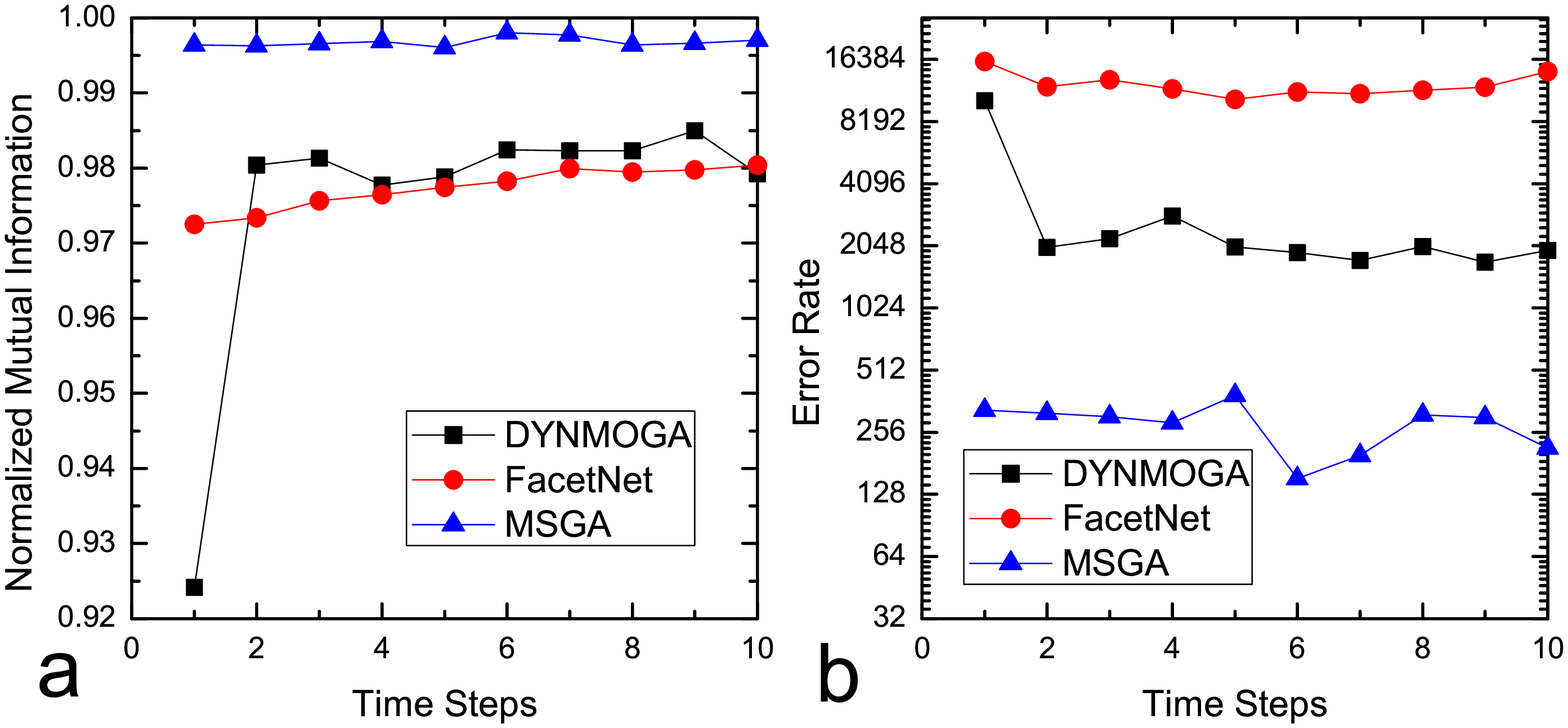}
\caption{Performances of the algorithms on intermittent communities network over 10 time steps.}
\label{fig:hide}
\end{figure}

Figure~\ref{fig:hide} illustrates the performance comparison among these algorithms on intermittent network. Clearly, MSGA exhibited a better performance than DYNMOGA and FacetNet at any time step in terms of both NMI and  error rate.

\subsection*{Real-world Network: Enron Email Network}
To test the performance of MSGA on real-world network, we applied this algorithm on the dataset of an email network. The network was established based on the communication record more than 150 senior employees of Enron corporation over three years~(from 1999 to 2002). The original dataset involves about 517431 emails owned by 151 users. We employed the version reduced by Folino et al~\cite{Folino14}, which only includes internal emails exchanged among the employees. We concentrated on analyzing the dataset on 2001 as Folino et al did, which contains the maximum number of emails.

At the beginning, an overall network $G_0$ was first constructed according to the communication record occurred in 2001 in the following way: a node represented an employee and a link was establish between a pair of nodes when the corresponding employees exchanged a mail in this year. Next, the communication record was divided into 12 subsets on the basis of the occurrence time, one for each month. And each subset was use to build a network as above-mentioned. In this way, we obtained 12 snapshots of a dynamic network, $G_1$,$G_2$,$\ldots$,$G_{12}$. To evaluate the accuracy of an algorithm at any time step, the ground truth should be known advanced. However, in practice the truth partition at each time step is unknown. We followed the same strategy of Lin et al~\cite{lin09} in which a best identified partition on the overall network $G_0$ is viewed as the benchmark partition for any time step. We applied MSGA on the $G_0$ and obtained a best partition including 14 communities.

Table~\ref{tab:email} lists the statistics of characteristics of each snapshot and information about the best partitions obtained by MSGA and DYNMOGA respectively~\footnote{It is notable that each snapshot tested was reduced by removing self links and isolated nodes with a self link.}. $T$ denotes a time step, $\left|E\right|$ indicates the number of weighted links, $\left|E^*\right|$ represents the number of different links, $Z$ is the average of the node, $CC$ is clustering coefficient. $Q$ and $\left|C\right|_D$ are modularity and the number of communities in a best partition identified by DYNMOGA respectively, while $TAS$ and $\left|E\right|_M$ indicate the temporal asymptotic surprise and the number of communities obtained by MSGA.
\begin{table*}
\centering
\caption{Statistics and community partition on email network at each time step}
\label{tab:email}
\begin{tabular}{lrcrcccccc}
\hline
             &      &       &       &       &      & \multicolumn{2}{c}{DYNMOGA} &\multicolumn{2}{c}{MSGA} \\
             \cline{7-8} \cline{9-10}
    T  &$\left|V\right|$  & $\left|E\right|$    & $\left|E^*\right|$    & Z     & $CC$   & $Q$    & $\left|C\right|_D$ & $TAS$     &$\left|E\right|_M$  \\
\hline
    1   & 95   & 988 & 166 & 3.4947 &0.2927 &0.5637    & 12   & 188.896 &16      \\
   2   & 92  & 1418 & 190 & 4.1304 & 0.4466 &0.5761   & 8  & 190.658 &14  \\
  3 & 94  & 1723 & 199 & 4.2340 & 0.3707 &0.5076   & 12  &181.11 & 13   \\
    4    &107  & 1691 & 240 & 4.486 & 0.3602 &0.5332 & 11 & 244.199 &19   \\
    5   &123 & 1718 & 272 & 4.4228 &  0.4181   &0.4827  & 12 & 250.551 &23  \\
    6  &120 & 864    &218 & 3.6333 & 0.3068    &0.6013      & 11 & 281.117 &20  \\
     7   & 108   & 1219 & 240 & 4.4444 &0.4158 &0.5772   & 10   & 248.025 &16      \\
    8    & 130 & 2063 & 371 & 5.7077 & 0.4011 &0.4646    & 11  & 314.223 &21  \\
    9 & 128  & 2967 & 342 & 5.3438& 0.0.4293 &0.5110  & 9  &329.668 &21   \\
    10    &133  & 8143 & 531 & 7.985 & 0.4696 &0.4191  & 11 & 376.546 &16   \\
    11    &127& 5861 & 438 & 6.8976 &  0.4623    &0.4430      & 10 & 346.673 &17  \\
    12  &113 & 1944     &306 & 5.4159 & 0.3919    &0.5105      & 9 & 304.025 &20  \\
\hline
\end{tabular}

\end{table*}

As we can see, the email network varies the number of nodes and links at each time step, coupled with the variations of average degree and clustering coefficient. As expected, the partitions identified by MSGA and DYNMOGA evolve over the time. At each time step DYNMOGA identified a partition with high modularity, implying that the network is with strong community structure for each snapshot in 2001. Apart from step 3, there exist significant differences between the number of communities that MSGA and DYNMOGA identified for other steps. This is mainly because one objective for DYNMOGA to optimize and the selector for the best solution from the Pareto front is modularity which has the resolution limit.

\begin{figure*}[ht]
\centering
\includegraphics[scale=0.5]{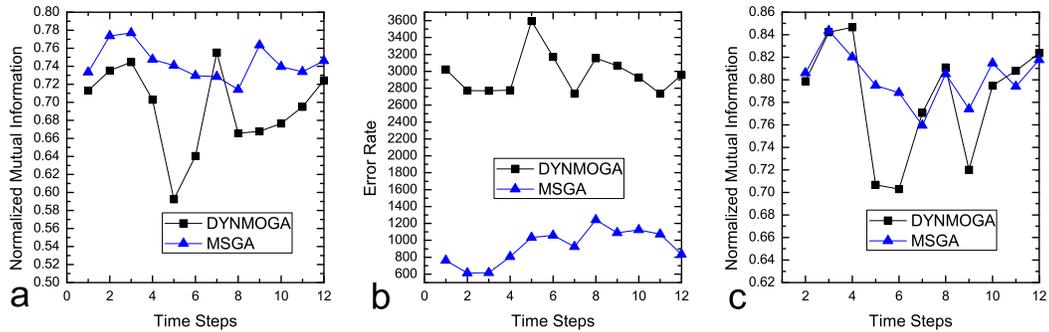}
\caption{Performance on Enron mail network (a)NMI (b) error rate (c)mutual information between adjacent partitions.}
\label{fig:email}
\end{figure*}

Figure~\ref{fig:email}~(a) and (b) reports the average performance in terms of NMI and error rate of MSGA and DYNMOGA, repectively. Compared with DYNMOGA, MSGA has a higher NMI apart from time step 7, and has a lower error rate at all time steps. Furthermore, figure~\ref{fig:email}(c) shows the mutual information between adjacent partitions identified by MSGA and DYNMOGA. It can be observed that both MSGA and DYNMOGA exhibit good agreement between partitions on adjacent time steps. MSGA shows a bit strong stability over all the time steps because for DYNMOGA the values at time steps 5, 6, 9 are significantly lower than those at other time steps.

To further gain insights into community evolution in depth, we analyzed the community structure of snapshots at time steps 1--3. Table~\ref{tab:partitions} lists the best partitions identified by the MSGA at the first and second time steps. It can be easily observed that several types of community events occurred at these time steps.

\begin{table*}[]
\centering
\caption{Community structure identified by MSGA at time steps 1 and 2.}
\label{tab:partitions}
\begin{tabular}{|c|l|c|l|}
\hline
\multicolumn{2}{|c|}{Time Step: 1}                     & \multicolumn{2}{c|}{Time Step: 2}                     \\ \hline
Comm. No. & Members                      & Comm. No. & Members                     \\ \hline
1             & 1 21 32 54 66 69 73 74 123 139 140 151 & 1             & 1 69 73 123 129 139 151               \\ \hline
2             & 2 19 28 70 141                         & 2             & 2 3 4 18 19 28 29 32 49 68 70 137 141 \\ \hline
3           & 3 4 18 29 68 137 & 3   & 5 37 89 \\ \hline
4 & 5 89 & 4  & 9 13 27 48 50 52 57 67 147 \\ \hline
5 & 9 48 50 57 67 147& 5 & 11 38 75 110 \\ \hline
6 & 11 75 & 6 & 15 86 93 97 115 130 132 133 140 148 149\\ \hline
7 & 13 27 52 & 7 & 17 21 25 26 40 59 61 71 90 125 146\\ \hline
8 & 17 25 26 40 49 56 59 61 71 91 100 125 128 145 146 & 8 &33 76 136 142 \\ \hline
9 & 33 58 76 136 142 & 9 &44 91 100 102 104 11113 27 52 \\ \hline
10 & 34 35 108 124 & 10 & 45 54 78 107 122 150\\ \hline
11 & 36 37 38 53 90 & 11 & 47 62 63 81 96 114 128 \\ \hline
12 & 42 44 47 62 84 96 102 104 111 & 12 & 53 58 77 118 \\ \hline
13 & 45 78 107 122 150 & 13 & 85 87 131 135 138\\ \hline
14 & 63 81 110 114 & 14 & 101 145  \\ \hline
15 & 85 138 &          &\\ \hline
16 & 86 87 93 97 115 130 132 133 135 149   &          & \\ \hline
\end{tabular}
\end{table*}

\begin{itemize}
\item Community contraction: for instance, the community $C_1^{t_1}$=\{\textbf{1} 21 32 54 66 \textbf{69 73} 74 \textbf{123 139} 140 \textbf{151}\} contracted to the community $C_1^{t_2}$:~\{1 69 73 123 129 139 151\} owing to the movement of the nodes 21, 32, 54, and 140 to other different communities and the disappearance of the nodes 66 and 74 at the seconde time step.
\item Community expansion: typically, community expansion was trigged by the participation of the existing nodes from other communities or~(and) new emerging nodes. For instance, the community $C_4^{t_1}$:~\{5 89\}  expanded to the community $C_4^{t_1}$:~\{5 \textbf{37} 89\} with the participation of the existing nodes 37 in the Community $C_{3}^{t_2}$.
\item Community merging: the community $C_2^{t_1}$:~\{\emph{2 19 28 70 141} \} and $C_3^{t_1}$:~\{\textbf{3 4 18 29 68 137 } \} are merged into a larger community $C_3^{t_2}$:~\{\emph{2}\textbf{ 3 4 18} \emph{19 28} \textbf{29} 32 49 \textbf{68} \emph{70} \textbf{137} \emph{141}\}, where nodes 32 and 49 joined from other communities.
\item Community death: it can be observed that there existed two possible cases for such type of event. The first case is that most of nodes in a community disappeared together and the remaining moved to other communities. For instance, the isolated community $C_9^{t_1}$:~\{34 35 108 124\} entirely disappeared at the second time step. The seconde case is that the community dissolved owing to the member moving to other different communities or disappearing. An typical example of this is the community $C_{11}^{t_1}$:~\{36 37 38 53 90\}.
\end{itemize}

\begin{figure*}[ht]
\centering
\includegraphics[scale=0.30]{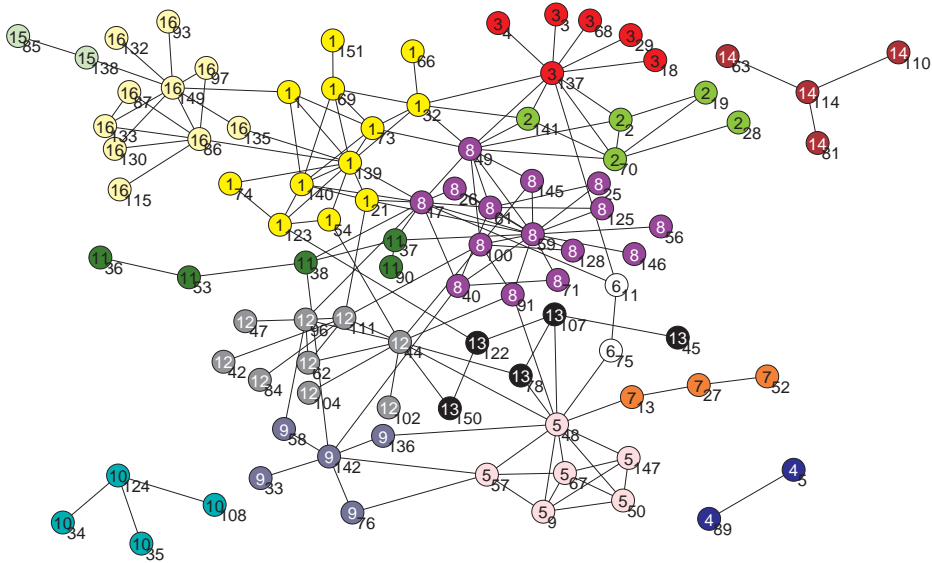}
\caption{Email network snapshot and the best partition at 1 time step. For each node, the centering number is the community label and the neighboring number is the node label.}
\label{fig:enron01}
\end{figure*}

\begin{figure*}[ht]
\centering
\includegraphics[scale=0.30]{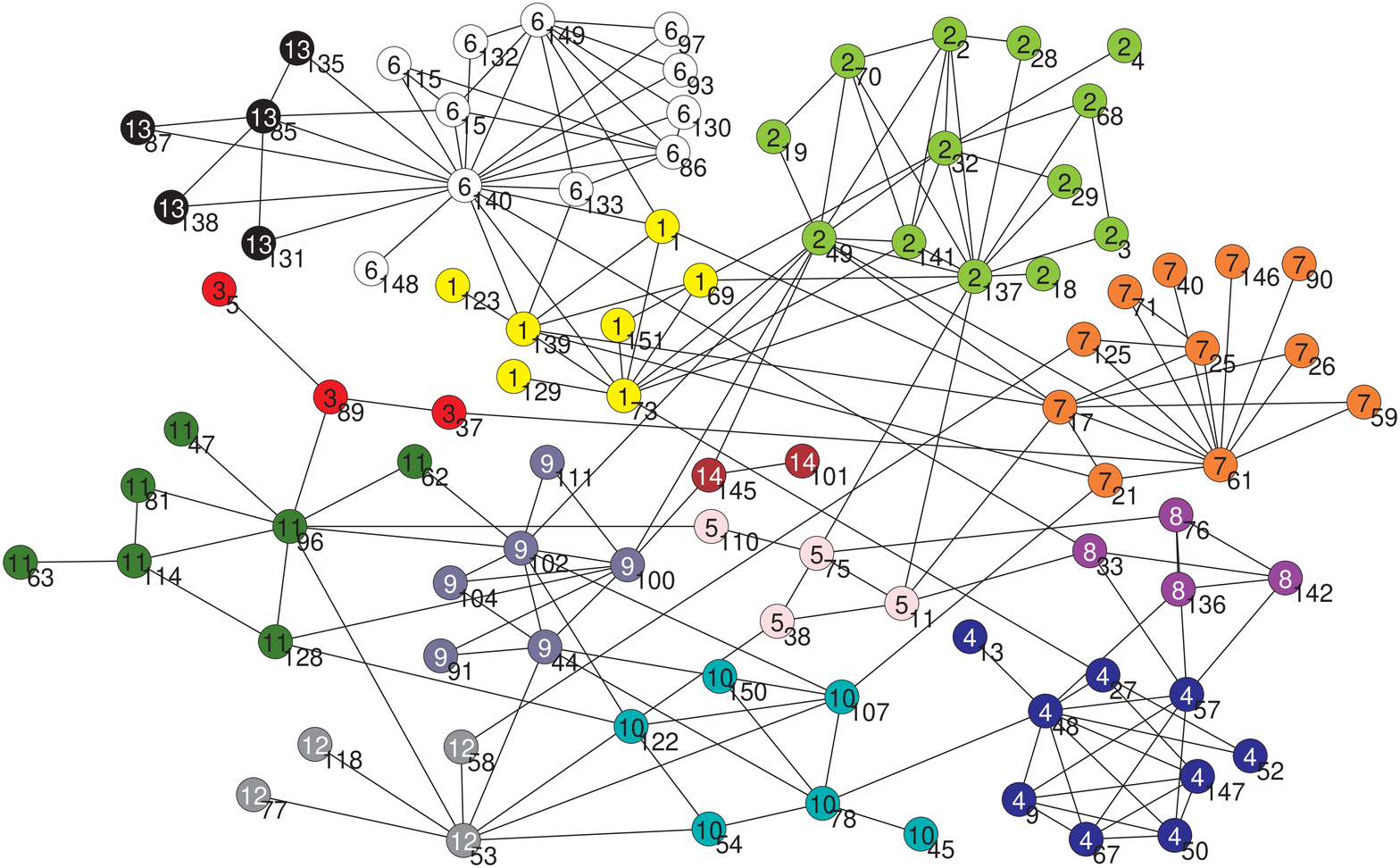}
\caption{Email network snapshot and the best partition at 2 time step. For each node, the centering number is the community label and the neighboring number is the node label.}
\label{fig:enron02}
\end{figure*}

\begin{figure*}[ht]
\centering
\includegraphics[scale=0.30]{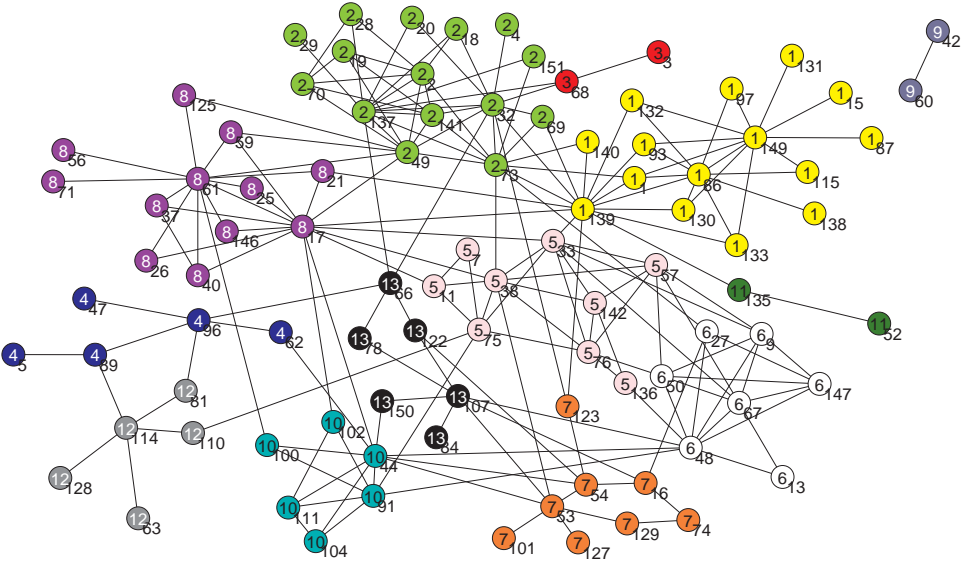}
\caption{Email network snapshot and the best partition at 3 time step. For each node, the centering number is the community label and the neighboring number is the node label.}
\label{fig:enron03}
\end{figure*}

Moreover, figure~\ref{fig:enron01}--\ref{fig:enron03} visualized the best partitions identified by the MSGA on the first three time steps. As we can see, at these time steps the MSGA indeed identified the community structure accurately while preserving good consistence in adjacent time steps.

\section{Conclusion}
Dynamic community detection is important for analyzing and predicting evolution of networked systems, as various dynamics has been discovered to be tightly correlated with community structure. In this paper, we presented an effective method for dynamic community detection using genetic algorithm. We have introduced a cost function called as temporal asymptotic surprise, TAS, that is used as the fitness function to optimize. TAS can be used to precisely reveal dynamic community structure with temporal smoothness and free of resolution limit. To effectively optimize TAS, we developed ad hoc merging and splitting operators for large-scale searching which are coupled with crossover and operation to find a better partition for each snapshot of a dynamic network.

Experimental results on synthetic networks and a real-world network show the method can more accurately reveal dynamic community structure than DYNMOGA and FacetNet. It also may suggest that temporal asymptotic surprise be a good cost function for dynamic community detection, and developing ad hoc merging and splitting operators with low cost be of benefit to the effectiveness of an evolution approach for community detection.

In summary, this method has three merits: It is free from the resolution limit; The adjacent community structures identified are temporal smoothness; There is no need to specify the number of communities in advance. We are looking forward to this method being applied to analyze the evolution of various real-world networks in the future.

\section*{References}
\bibliography{dynamic}

\begin{thebibliography}{47}
\expandafter\ifx\csname natexlab\endcsname\relax\def\natexlab#1{#1}\fi
\expandafter\ifx\csname bibnamefont\endcsname\relax
  \def\bibnamefont#1{#1}\fi
\expandafter\ifx\csname bibfnamefont\endcsname\relax
  \def\bibfnamefont#1{#1}\fi
\expandafter\ifx\csname citenamefont\endcsname\relax
  \def\citenamefont#1{#1}\fi
\expandafter\ifx\csname url\endcsname\relax
  \def\url#1{\texttt{#1}}\fi
\expandafter\ifx\csname urlprefix\endcsname\relax\def\urlprefix{URL }\fi
\providecommand{\bibinfo}[2]{#2}
\providecommand{\eprint}[2][]{\url{#2}}

\bibitem[{\citenamefont{Erkan and Radev}(2004)}]{Erkan04}
\bibinfo{author}{\bibfnamefont{G.}~\bibnamefont{Erkan}} \bibnamefont{and}
  \bibinfo{author}{\bibfnamefont{D.~R.} \bibnamefont{Radev}},
  \bibinfo{journal}{J. Artif. Int. Res.} \textbf{\bibinfo{volume}{22}},
  \bibinfo{pages}{457} (\bibinfo{year}{2004}).

\bibitem[{\citenamefont{V\'eronis}(2004)}]{veronis04}
\bibinfo{author}{\bibfnamefont{J.}~\bibnamefont{V\'eronis}},
  \bibinfo{journal}{Computer Speech and Language}
  \textbf{\bibinfo{volume}{18}}, \bibinfo{pages}{223} (\bibinfo{year}{2004}).

\bibitem[{\citenamefont{Lv et~al.}(2002)\citenamefont{Lv, Cao, Cohen, Li, and
  Shenker}}]{Lv02}
\bibinfo{author}{\bibfnamefont{Q.}~\bibnamefont{Lv}},
  \bibinfo{author}{\bibfnamefont{P.}~\bibnamefont{Cao}},
  \bibinfo{author}{\bibfnamefont{E.}~\bibnamefont{Cohen}},
  \bibinfo{author}{\bibfnamefont{K.}~\bibnamefont{Li}}, \bibnamefont{and}
  \bibinfo{author}{\bibfnamefont{S.}~\bibnamefont{Shenker}}, in
  \emph{\bibinfo{booktitle}{Proceedings of the 16th International Conference on
  Supercomputing}} (\bibinfo{publisher}{ACM}, \bibinfo{address}{New York, NY,
  USA}, \bibinfo{year}{2002}), ICS '02, pp. \bibinfo{pages}{84--95}.

\bibitem[{\citenamefont{Zhang et~al.}(2017)\citenamefont{Zhang, Zhang, Fan,
  Tang, and Deng}}]{Zhang17}
\bibinfo{author}{\bibfnamefont{Z.-P.} \bibnamefont{Zhang}},
  \bibinfo{author}{\bibfnamefont{J.-P.} \bibnamefont{Zhang}},
  \bibinfo{author}{\bibfnamefont{C.}~\bibnamefont{Fan}},
  \bibinfo{author}{\bibfnamefont{Y.-J.} \bibnamefont{Tang}}, \bibnamefont{and}
  \bibinfo{author}{\bibfnamefont{L.}~\bibnamefont{Deng}},
  \bibinfo{journal}{IEEE/ACM Transactions on Computational Biology and
  Bioinformatics}  (\bibinfo{year}{2017}).

\bibitem[{\citenamefont{Xiao et~al.}(2017)\citenamefont{Xiao, Zhang, and
  Deng}}]{Xiao17}
\bibinfo{author}{\bibfnamefont{Y.}~\bibnamefont{Xiao}},
  \bibinfo{author}{\bibfnamefont{J.-P.} \bibnamefont{Zhang}}, \bibnamefont{and}
  \bibinfo{author}{\bibfnamefont{L.}~\bibnamefont{Deng}},
  \bibinfo{journal}{Sci. Rep.} \textbf{\bibinfo{volume}{7}},
  \bibinfo{pages}{3664} (\bibinfo{year}{2017}).

\bibitem[{\citenamefont{Watts and Strogatz}(1998)}]{Watts98}
\bibinfo{author}{\bibfnamefont{D.~J.} \bibnamefont{Watts}} \bibnamefont{and}
  \bibinfo{author}{\bibfnamefont{S.~H.} \bibnamefont{Strogatz}},
  \bibinfo{journal}{Nature} \textbf{\bibinfo{volume}{393}},
  \bibinfo{pages}{440} (\bibinfo{year}{1998}).

\bibitem[{\citenamefont{Barab\'asi and Albert}(1999)}]{Barabasi99}
\bibinfo{author}{\bibfnamefont{A.-L.} \bibnamefont{Barab\'asi}}
  \bibnamefont{and} \bibinfo{author}{\bibfnamefont{R.}~\bibnamefont{Albert}},
  \bibinfo{journal}{Science} \textbf{\bibinfo{volume}{286}},
  \bibinfo{pages}{509} (\bibinfo{year}{1999}).

\bibitem[{\citenamefont{Newman}(2003)}]{Newman03}
\bibinfo{author}{\bibfnamefont{M.~E.~J.} \bibnamefont{Newman}},
  \bibinfo{journal}{SIAM Review} \textbf{\bibinfo{volume}{45}},
  \bibinfo{pages}{167} (\bibinfo{year}{2003}).

\bibitem[{\citenamefont{Girvan and Newman}(2002)}]{Girvan:2002a}
\bibinfo{author}{\bibfnamefont{M.}~\bibnamefont{Girvan}} \bibnamefont{and}
  \bibinfo{author}{\bibfnamefont{M.~E.~J.} \bibnamefont{Newman}},
  \bibinfo{journal}{PNAS} \textbf{\bibinfo{volume}{99}}, \bibinfo{pages}{7821}
  (\bibinfo{year}{2002}).

\bibitem[{\citenamefont{Newman}(2006)}]{Newman06a}
\bibinfo{author}{\bibfnamefont{M.~E.~J.} \bibnamefont{Newman}},
  \bibinfo{journal}{Proc. Natl. Acad. Sci. USA} \textbf{\bibinfo{volume}{103}},
  \bibinfo{pages}{8577} (\bibinfo{year}{2006}).

\bibitem[{\citenamefont{Newman}(2013)}]{Newman13}
\bibinfo{author}{\bibfnamefont{M.~E.~J.} \bibnamefont{Newman}},
  \bibinfo{journal}{Phys. Rev. E} \textbf{\bibinfo{volume}{88}},
  \bibinfo{pages}{042822} (\bibinfo{year}{2013}).

\bibitem[{\citenamefont{Guimer\`a and Amaral}(2005)}]{Guimera05}
\bibinfo{author}{\bibfnamefont{R.}~\bibnamefont{Guimer\`a}} \bibnamefont{and}
  \bibinfo{author}{\bibfnamefont{L.~A.~N.} \bibnamefont{Amaral}},
  \bibinfo{journal}{Nature} \textbf{\bibinfo{volume}{433}},
  \bibinfo{pages}{895} (\bibinfo{year}{2005}).

\bibitem[{\citenamefont{Duch and Arenas}(2005)}]{Duch05}
\bibinfo{author}{\bibfnamefont{J.}~\bibnamefont{Duch}} \bibnamefont{and}
  \bibinfo{author}{\bibfnamefont{A.}~\bibnamefont{Arenas}},
  \bibinfo{journal}{Phys. Rev. E} \textbf{\bibinfo{volume}{72}},
  \bibinfo{pages}{027104} (\bibinfo{year}{2005}).

\bibitem[{\citenamefont{Raghavan et~al.}(2007)\citenamefont{Raghavan, Albert,
  and Kumara}}]{Raghavan07}
\bibinfo{author}{\bibfnamefont{U.~N.} \bibnamefont{Raghavan}},
  \bibinfo{author}{\bibfnamefont{R.}~\bibnamefont{Albert}}, \bibnamefont{and}
  \bibinfo{author}{\bibfnamefont{S.}~\bibnamefont{Kumara}},
  \bibinfo{journal}{Phys. Rev. E} \textbf{\bibinfo{volume}{76}},
  \bibinfo{pages}{036106} (\bibinfo{year}{2007}).

\bibitem[{\citenamefont{Zhang et~al.}(2007)\citenamefont{Zhang, Wang, and
  Zhang}}]{Zhang07a}
\bibinfo{author}{\bibfnamefont{S.-H.} \bibnamefont{Zhang}},
  \bibinfo{author}{\bibfnamefont{R.-S.} \bibnamefont{Wang}}, \bibnamefont{and}
  \bibinfo{author}{\bibfnamefont{X.-S.} \bibnamefont{Zhang}},
  \bibinfo{journal}{Phys. Rev. E} \textbf{\bibinfo{volume}{76}},
  \bibinfo{pages}{046103} (\bibinfo{year}{2007}).

\bibitem[{\citenamefont{Rosvall and Bergstrom}(2007)}]{Rosvall07}
\bibinfo{author}{\bibfnamefont{M.}~\bibnamefont{Rosvall}} \bibnamefont{and}
  \bibinfo{author}{\bibfnamefont{C.}~\bibnamefont{Bergstrom}},
  \bibinfo{journal}{Proc. Natl. Acad. Sci. USA} \textbf{\bibinfo{volume}{104}},
  \bibinfo{pages}{7327} (\bibinfo{year}{2007}).

\bibitem[{\citenamefont{Rosvall and Bergstrom}(2008)}]{Rosvall08}
\bibinfo{author}{\bibfnamefont{M.}~\bibnamefont{Rosvall}} \bibnamefont{and}
  \bibinfo{author}{\bibfnamefont{C.}~\bibnamefont{Bergstrom}},
  \bibinfo{journal}{Proc. Natl. Acad. Sci. USA} \textbf{\bibinfo{volume}{105}},
  \bibinfo{pages}{1118} (\bibinfo{year}{2008}).

\bibitem[{\citenamefont{Hastings}(2006)}]{Hasting06}
\bibinfo{author}{\bibfnamefont{M.~B.} \bibnamefont{Hastings}},
  \bibinfo{journal}{hys. Rev. E} \textbf{\bibinfo{volume}{74}},
  \bibinfo{pages}{035102} (\bibinfo{year}{2006}).

\bibitem[{\citenamefont{Newman and Leicht}(2007)}]{Newman07}
\bibinfo{author}{\bibfnamefont{M.~E.~J.} \bibnamefont{Newman}}
  \bibnamefont{and} \bibinfo{author}{\bibfnamefont{E.~A.}
  \bibnamefont{Leicht}}, \bibinfo{journal}{Proc. Natl. Acad. Sci. USA}
  \textbf{\bibinfo{volume}{104}}, \bibinfo{pages}{9564} (\bibinfo{year}{2007}).

\bibitem[{\citenamefont{Fortunato}(2010)}]{Fortunato10}
\bibinfo{author}{\bibfnamefont{S.}~\bibnamefont{Fortunato}},
  \bibinfo{journal}{Phys. Rep.} \textbf{\bibinfo{volume}{486}},
  \bibinfo{pages}{75} (\bibinfo{year}{2010}).

\bibitem[{\citenamefont{Good et~al.}(2010)\citenamefont{Good, de~Montjoye, and
  Clauset}}]{Good10}
\bibinfo{author}{\bibfnamefont{B.~H.} \bibnamefont{Good}},
  \bibinfo{author}{\bibfnamefont{Y.-A.} \bibnamefont{de~Montjoye}},
  \bibnamefont{and} \bibinfo{author}{\bibfnamefont{A.}~\bibnamefont{Clauset}},
  \bibinfo{journal}{Phys. Rev. E} \textbf{\bibinfo{volume}{81}},
  \bibinfo{pages}{046106} (\bibinfo{year}{2010}).

\bibitem[{\citenamefont{Chakrabarti et~al.}(2006)\citenamefont{Chakrabarti,
  Kumar, and Tomkins}}]{Chakrabarti06}
\bibinfo{author}{\bibfnamefont{D.}~\bibnamefont{Chakrabarti}},
  \bibinfo{author}{\bibfnamefont{R.}~\bibnamefont{Kumar}}, \bibnamefont{and}
  \bibinfo{author}{\bibfnamefont{A.}~\bibnamefont{Tomkins}}, in
  \emph{\bibinfo{booktitle}{the 12th ACM SIGKDD International Conference on
  Knowledge Discovery and Data Mining}} (\bibinfo{publisher}{ACM},
  \bibinfo{year}{2006}), pp. \bibinfo{pages}{554--560}.

\bibitem[{\citenamefont{Lin et~al.}(2009)\citenamefont{Lin, Zhu, Sundaram, and
  Tseng}}]{lin09}
\bibinfo{author}{\bibfnamefont{Y.-R.} \bibnamefont{Lin}},
  \bibinfo{author}{\bibfnamefont{S.}~\bibnamefont{Zhu}},
  \bibinfo{author}{\bibfnamefont{H.}~\bibnamefont{Sundaram}}, \bibnamefont{and}
  \bibinfo{author}{\bibfnamefont{B.~L.} \bibnamefont{Tseng}},
  \bibinfo{journal}{ACM Transaction on Knowledge Discovery from Data}
  \textbf{\bibinfo{volume}{3}}, \bibinfo{pages}{1} (\bibinfo{year}{2009}).

\bibitem[{\citenamefont{Chi et~al.}(2009)\citenamefont{Chi, Song, Zhou, Hino,
  and Tseng}}]{Chi09}
\bibinfo{author}{\bibfnamefont{Y.}~\bibnamefont{Chi}},
  \bibinfo{author}{\bibfnamefont{X.}~\bibnamefont{Song}},
  \bibinfo{author}{\bibfnamefont{D.}~\bibnamefont{Zhou}},
  \bibinfo{author}{\bibfnamefont{K.}~\bibnamefont{Hino}}, \bibnamefont{and}
  \bibinfo{author}{\bibfnamefont{B.-L.} \bibnamefont{Tseng}},
  \bibinfo{journal}{ACM Trans. Knowl. Discov. Data}
  \textbf{\bibinfo{volume}{3}}, \bibinfo{pages}{17:1} (\bibinfo{year}{2009}).

\bibitem[{\citenamefont{Kim and Han}(2009)}]{Kim09}
\bibinfo{author}{\bibfnamefont{M.-S.} \bibnamefont{Kim}} \bibnamefont{and}
  \bibinfo{author}{\bibfnamefont{J.~A.} \bibnamefont{Han}},
  \bibinfo{journal}{PVLDB} \textbf{\bibinfo{volume}{2}}, \bibinfo{pages}{622}
  (\bibinfo{year}{2009}).

\bibitem[{\citenamefont{Folino and Pizzuti}(2014)}]{Folino14}
\bibinfo{author}{\bibfnamefont{F.}~\bibnamefont{Folino}} \bibnamefont{and}
  \bibinfo{author}{\bibfnamefont{C.}~\bibnamefont{Pizzuti}},
  \bibinfo{journal}{IEEE Transaction on Knowledge and Data Engineering}
  \textbf{\bibinfo{volume}{26}}, \bibinfo{pages}{1838} (\bibinfo{year}{2014}).

\bibitem[{\citenamefont{Qin et~al.}(2016)\citenamefont{Qin, Dai, Jiao, Wang,
  and Yuan}}]{Qin16}
\bibinfo{author}{\bibfnamefont{X.-M.} \bibnamefont{Qin}},
  \bibinfo{author}{\bibfnamefont{W.-D.} \bibnamefont{Dai}},
  \bibinfo{author}{\bibfnamefont{P.-F.} \bibnamefont{Jiao}},
  \bibinfo{author}{\bibfnamefont{W.-J.} \bibnamefont{Wang}}, \bibnamefont{and}
  \bibinfo{author}{\bibfnamefont{N.}~\bibnamefont{Yuan}},
  \bibinfo{journal}{Sci. Rep.} \textbf{\bibinfo{volume}{6}},
  \bibinfo{pages}{31454} (\bibinfo{year}{2016}).

\bibitem[{\citenamefont{Wang et~al.}(2017)\citenamefont{Wang, Gao, and
  Ma}}]{Wang17}
\bibinfo{author}{\bibfnamefont{P.-Z.} \bibnamefont{Wang}},
  \bibinfo{author}{\bibfnamefont{L.}~\bibnamefont{Gao}}, \bibnamefont{and}
  \bibinfo{author}{\bibfnamefont{X.-K.} \bibnamefont{Ma}}, \bibinfo{journal}{J.
  Stat. Mech.} p. \bibinfo{pages}{P013401} (\bibinfo{year}{2017}).

\bibitem[{\citenamefont{Gong et~al.}(2012)\citenamefont{Gong, Zhang, Ma, and
  Jiao}}]{Gong12}
\bibinfo{author}{\bibfnamefont{M.-G.} \bibnamefont{Gong}},
  \bibinfo{author}{\bibfnamefont{L.-J.} \bibnamefont{Zhang}},
  \bibinfo{author}{\bibfnamefont{J.-J.} \bibnamefont{Ma}}, \bibnamefont{and}
  \bibinfo{author}{\bibfnamefont{L.-C.} \bibnamefont{Jiao}},
  \bibinfo{journal}{Journal of Computer science and technology}
  \textbf{\bibinfo{volume}{27}}, \bibinfo{pages}{455} (\bibinfo{year}{2012}).

\bibitem[{\citenamefont{Kannan et~al.}(2004)\citenamefont{Kannan, Vempala, and
  Vetta}}]{Kannan04}
\bibinfo{author}{\bibfnamefont{R.}~\bibnamefont{Kannan}},
  \bibinfo{author}{\bibfnamefont{S.}~\bibnamefont{Vempala}}, \bibnamefont{and}
  \bibinfo{author}{\bibfnamefont{A.}~\bibnamefont{Vetta}},
  \bibinfo{journal}{Journal of the ACM} \textbf{\bibinfo{volume}{51}},
  \bibinfo{pages}{497} (\bibinfo{year}{2004}).

\bibitem[{\citenamefont{Shi and Malik}(2000)}]{Shi00}
\bibinfo{author}{\bibfnamefont{J.}~\bibnamefont{Shi}} \bibnamefont{and}
  \bibinfo{author}{\bibfnamefont{J.}~\bibnamefont{Malik}},
  \bibinfo{journal}{IEEE Transaction on Pattern analysis and machine
  intelligence} \textbf{\bibinfo{volume}{22}}, \bibinfo{pages}{888}
  (\bibinfo{year}{2000}).

\bibitem[{\citenamefont{Folino and Pizzuti}(2010)}]{Folino10}
\bibinfo{author}{\bibfnamefont{F.}~\bibnamefont{Folino}} \bibnamefont{and}
  \bibinfo{author}{\bibfnamefont{C.}~\bibnamefont{Pizzuti}}, in
  \emph{\bibinfo{booktitle}{International Conference on advances in social
  networks analysis and mining~(ASONAM10)}} (\bibinfo{year}{2010}), pp.
  \bibinfo{pages}{256--263}.

\bibitem[{\citenamefont{Traag et~al.}(2015)\citenamefont{Traag, Aldecoa, and
  Delvenne}}]{Traag15}
\bibinfo{author}{\bibfnamefont{V.~A.} \bibnamefont{Traag}},
  \bibinfo{author}{\bibfnamefont{R.}~\bibnamefont{Aldecoa}}, \bibnamefont{and}
  \bibinfo{author}{\bibfnamefont{J.-C.} \bibnamefont{Delvenne}},
  \bibinfo{journal}{Phys. Rev. E} \textbf{\bibinfo{volume}{92}},
  \bibinfo{pages}{022816} (\bibinfo{year}{2015}).

\bibitem[{\citenamefont{Newman and Girvan}(2004)}]{Newman04a}
\bibinfo{author}{\bibfnamefont{M.~E.~J.} \bibnamefont{Newman}}
  \bibnamefont{and} \bibinfo{author}{\bibfnamefont{M.}~\bibnamefont{Girvan}},
  \bibinfo{journal}{Phys. Rev. E} \textbf{\bibinfo{volume}{69}},
  \bibinfo{pages}{026113} (\bibinfo{year}{2004}).

\bibitem[{\citenamefont{Fortunato and Barthelemy}(2007)}]{Fortunato07}
\bibinfo{author}{\bibfnamefont{S.}~\bibnamefont{Fortunato}} \bibnamefont{and}
  \bibinfo{author}{\bibfnamefont{M.}~\bibnamefont{Barthelemy}},
  \bibinfo{journal}{Proc. Nat. Acad. Sci. USA} \textbf{\bibinfo{volume}{104}},
  \bibinfo{pages}{36} (\bibinfo{year}{2007}).

\bibitem[{\citenamefont{Li et~al.}(2008)\citenamefont{Li, Zhang, Wang, Zhang,
  and Chen}}]{Li08}
\bibinfo{author}{\bibfnamefont{Z.-P.} \bibnamefont{Li}},
  \bibinfo{author}{\bibfnamefont{S.-H.} \bibnamefont{Zhang}},
  \bibinfo{author}{\bibfnamefont{R.-S.} \bibnamefont{Wang}},
  \bibinfo{author}{\bibfnamefont{X.-S.} \bibnamefont{Zhang}}, \bibnamefont{and}
  \bibinfo{author}{\bibfnamefont{L.-N.} \bibnamefont{Chen}},
  \bibinfo{journal}{Phys. Rev. E} \textbf{\bibinfo{volume}{77}},
  \bibinfo{pages}{019901} (\bibinfo{year}{2008}).

\bibitem[{\citenamefont{Ronhovde and Z.}(2010)}]{Ronhovde10}
\bibinfo{author}{\bibfnamefont{P.}~\bibnamefont{Ronhovde}} \bibnamefont{and}
  \bibinfo{author}{\bibfnamefont{N.}~\bibnamefont{Z.}}, \bibinfo{journal}{Phys.
  Rev. E} \textbf{\bibinfo{volume}{81}}, \bibinfo{pages}{046114}
  (\bibinfo{year}{2010}).

\bibitem[{\citenamefont{Aldecoa and Martin}(2011)}]{Aldecoa11}
\bibinfo{author}{\bibfnamefont{R.}~\bibnamefont{Aldecoa}} \bibnamefont{and}
  \bibinfo{author}{\bibfnamefont{I.}~\bibnamefont{Martin}},
  \bibinfo{journal}{PloS ONE} \textbf{\bibinfo{volume}{6}},
  \bibinfo{pages}{e24195} (\bibinfo{year}{2011}).

\bibitem[{\citenamefont{Nicolin and Bifone}(2016)}]{Nicolini16}
\bibinfo{author}{\bibfnamefont{C.}~\bibnamefont{Nicolin}} \bibnamefont{and}
  \bibinfo{author}{\bibfnamefont{A.}~\bibnamefont{Bifone}},
  \bibinfo{journal}{Sci. Rep.} p. \bibinfo{pages}{19250}
  (\bibinfo{year}{2016}).

\bibitem[{\citenamefont{Aldecoa and Martin}(2013)}]{Aldecoa13}
\bibinfo{author}{\bibfnamefont{R.}~\bibnamefont{Aldecoa}} \bibnamefont{and}
  \bibinfo{author}{\bibfnamefont{I.}~\bibnamefont{Martin}},
  \bibinfo{journal}{Sci. Rep.} p. \bibinfo{pages}{013401}
  (\bibinfo{year}{2013}).

\bibitem[{\citenamefont{Jiang et~al.}(2014)\citenamefont{Jiang, Jia, and
  Yu}}]{Jiang14}
\bibinfo{author}{\bibfnamefont{Y.-W.} \bibnamefont{Jiang}},
  \bibinfo{author}{\bibfnamefont{C.-Y.} \bibnamefont{Jia}}, \bibnamefont{and}
  \bibinfo{author}{\bibfnamefont{J.}~\bibnamefont{Yu}}, \bibinfo{journal}{J.
  Phys. A} \textbf{\bibinfo{volume}{47}}, \bibinfo{pages}{165101}
  (\bibinfo{year}{2014}).

\bibitem[{\citenamefont{Pizzuti}(2008)}]{Pizzuti08}
\bibinfo{author}{\bibfnamefont{C.}~\bibnamefont{Pizzuti}}, in
  \emph{\bibinfo{booktitle}{Lecture Notes in Computer Science}}, edited by
  \bibinfo{editor}{\bibfnamefont{G.}~\bibnamefont{Rudolph}}
  (\bibinfo{publisher}{Springer-Verlag}, \bibinfo{year}{2008}), vol.
  \bibinfo{volume}{5199}, pp. \bibinfo{pages}{1081--1090}.

\bibitem[{\citenamefont{Zhan et~al.}(2011)\citenamefont{Zhan, Zhang, Guan, and
  S.-G.}}]{zhan11}
\bibinfo{author}{\bibfnamefont{W.-H.} \bibnamefont{Zhan}},
  \bibinfo{author}{\bibfnamefont{Z.-Z.} \bibnamefont{Zhang}},
  \bibinfo{author}{\bibfnamefont{J.-H.} \bibnamefont{Guan}}, \bibnamefont{and}
  \bibinfo{author}{\bibfnamefont{Z.}~\bibnamefont{S.-G.}},
  \bibinfo{journal}{Phys. Rev. E} \textbf{\bibinfo{volume}{83}},
  \bibinfo{pages}{066120} (\bibinfo{year}{2011}).

\bibitem[{\citenamefont{Zhan et~al.}(2016)\citenamefont{Zhan, Guan, Chen, Niu,
  and G.}}]{zhan16}
\bibinfo{author}{\bibfnamefont{W.-H.} \bibnamefont{Zhan}},
  \bibinfo{author}{\bibfnamefont{J.-H.} \bibnamefont{Guan}},
  \bibinfo{author}{\bibfnamefont{H.-H.} \bibnamefont{Chen}},
  \bibinfo{author}{\bibfnamefont{J.}~\bibnamefont{Niu}}, \bibnamefont{and}
  \bibinfo{author}{\bibfnamefont{J.}~\bibnamefont{G.}},
  \bibinfo{journal}{Physica A} \textbf{\bibinfo{volume}{442}},
  \bibinfo{pages}{182} (\bibinfo{year}{2016}).

\bibitem[{\citenamefont{Danon et~al.}(2005)\citenamefont{Danon, D\'iaz-Guilera,
  Duch, and Arenas}}]{Danon05}
\bibinfo{author}{\bibfnamefont{L.}~\bibnamefont{Danon}},
  \bibinfo{author}{\bibfnamefont{A.}~\bibnamefont{D\'iaz-Guilera}},
  \bibinfo{author}{\bibfnamefont{J.}~\bibnamefont{Duch}}, \bibnamefont{and}
  \bibinfo{author}{\bibfnamefont{A.}~\bibnamefont{Arenas}},
  \bibinfo{journal}{J. Stat. Mech.} p. \bibinfo{pages}{P09008}
  (\bibinfo{year}{2005}).

\bibitem[{\citenamefont{Greene et~al.}(2010)\citenamefont{Greene, Doyle, and
  Cunningham}}]{Greene10}
\bibinfo{author}{\bibfnamefont{D.}~\bibnamefont{Greene}},
  \bibinfo{author}{\bibfnamefont{D.}~\bibnamefont{Doyle}}, \bibnamefont{and}
  \bibinfo{author}{\bibfnamefont{P.}~\bibnamefont{Cunningham}}, in
  \emph{\bibinfo{booktitle}{International Conference on Advances in Social
  Networks Analysis and Mining}} (\bibinfo{year}{2010}).

\bibitem[{\citenamefont{Lancichinetti and S.}(2008)}]{Lancichinetti08}
\bibinfo{author}{\bibfnamefont{A.}~\bibnamefont{Lancichinetti}}
  \bibnamefont{and} \bibinfo{author}{\bibfnamefont{F.}~\bibnamefont{S.}},
  \bibinfo{journal}{Phys. Rev. E} \textbf{\bibinfo{volume}{78}},
  \bibinfo{pages}{046110} (\bibinfo{year}{2008}).

\end{thebibliography}
\section*{Acknowledgment}
This work was supported by the Zhejiang Provincial Natural Science Foundation of China under Grant No.~LY13F020038,~LY15F020010, Zhejiang Public technology applied research projects under Grant No.2014C31059, and the open Project of the State Key Laboratory of Software Engineering under Grant No. SKLST2014-10-15.
\end{document}